\documentclass[twocolumn,pra]{revtex4-2}

\usepackage{graphicx}
\usepackage{dcolumn}
\usepackage{bm}
\usepackage{braket}
\usepackage{amsmath}
\usepackage{amssymb}
\usepackage{amsfonts}
\usepackage{bbm}
\usepackage{color}

\begin{document}
\preprint{APS/123-QED}
\title{Rényi entropy of the permutationally invariant part of the ground state across a quantum phase transition}

\author{Yuki Miyazaki$^1$}
\author{Giacomo Marmorini$^{1,2}$}
\author{Nobuo Furukawa$^1$}
\author{Daisuke Yamamoto$^2$}

\affiliation{$^1$Department of Physical Science, Aoyama Gakuin University, Kanagawa 252-5258, Japan}
\affiliation{$^2$Department of Physics, College of Humanities and Sciences, Nihon University, Tokyo 156-8550, Japan}

\date{\today}
\begin{abstract}
We investigate the role of the permutationally invariant part of the density matrix (PIDM) in capturing the properties of the ground state of the system during a quantum phase transition.
In the context of quantum state tomography, PIDM is known to be obtainable with only a low number of measurement settings, namely $\mathcal{O}(L^2)$, where $L$ is the system size.
Considering the transverse-field Ising chain as an example, we compute the second-order Rényi entropy of PIDM for the ground state by using the density matrix renormalization group algorithm.
In the ferromagnetic case, the ground state is permutationally invariant both in the limits of zero and infinite field, leading to vanishing Rényi entropy of PIDM.
The latter exhibits a broad peak as a function of the transverse field around the quantum critical point, which gets more pronounced for larger system size.
In the antiferromagnetic case, the peak structure disappears and the Rényi entropy diverges like $\mathcal{O}(L)$ in the whole field range of the ordered phase.
We discuss the cause of these behaviors of the Rényi entropy of PIDM, examining the possible application of this experimentally tractable quantity to the analysis of phase transition phenomena.  
\end{abstract}

\maketitle

\section{introduction}
Historically, in the field of solid state physics, macroscopic observables have been measured to detect a phase transition, e.g., magnetization for magnetic phase transitions \cite{MT_dimer1,MT_dimer2} and electrical resistance for superconductivity \cite{SC1,SC2}.
Recently topological phases such as quantum Hall effect \cite{topo1,topo2} and highly entangled phases including spin liquid phases in the Kitaev model \cite{Kitaev} have been attracting greater attention since they evade the paradigm of symmetry breaking and conventional order parameters.
Quantum information theory offers an insightful perspective on this issue.
For example, the so-called topological entanglement entropy, which appears in the von Neumann entanglement entropy as a universal constant, characterizes topologically ordered phases \cite{topo_ent1,topo_ent2};
in other cases, the structure of the entanglement spectrum helps us distinguish spin liquid phase from magnetically ordered phases \cite{ent_spect1,ent_spect2}.
In modern condensed matter physics, there is a growing demand for both theoretical analysis and experimental measurement of quantities related to quantum information.

In artificial quantum systems such as cold atomic and molecular gases in an optical lattice \cite{coldatom1,coldatom2,coldatom3,coldatom4}, photonic systems \cite{photonic1,photonic2}, trapped-ion systems \cite{trappedion1,trappedion2}, Rydberg atom arrays in optical tweezers \cite{Rydberg1,Rydberg2,Rydberg3,Rydberg4,Rydberg5},
and superconducting circuits \cite{SCcircuit},
the simulation of quantum many-body physics can be conducted in an ideal environment, taking advantage of its cleanness and high controllability.
Using these artificial quantum systems, the reconstruction of a density matrix has been conducted by preparing the identical state repeatedly and measuring the expectation value of various observables \cite{QST_exp_photon,QST_exp_iontrap,QST_exp_SC}.
The procedure to reconstruct a density matrix is known as quantum state tomography (QST) \cite{QST1,QST2,QST3,QST4}. 

\begin{figure}[t]
    \centering
    \includegraphics[width=6.5cm]{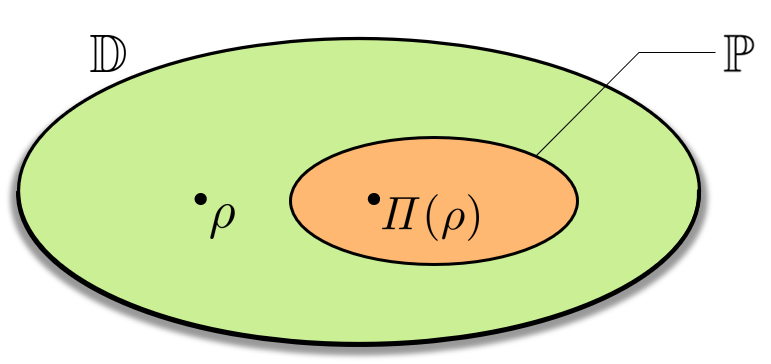}
    \caption{Schematic sketch of the linear spaces $\mathbb{D}$ and $\mathbb{P}$ spanned by the orthogonal operators $\{\bigotimes_{i=1}^L{\sigma}_{\kappa_i}\}$ and by the orthogonal PI operators $\{\mathcal{B}_d^{(n)}\}$ or the PI random operators $\{\mathcal{A}_d^{(n)}\}$, respectively. 
    Any density matrix $\rho$ is an element of the space $\mathbb{D}$.
    The PI part of an arbitrary density matrix of an $L$-qubit system, $\mathit{\Pi}(\rho)$, exist in $\mathbb{P}\subset\mathbb{D}$.}
   \label{space}
\end{figure}

In the QST for $L$-qubit system, an arbitrary density matrix $\rho$, which is a $2^L$ by $2^L$ matrix, can be expanded in the basis $\{\bigotimes_{i=1}^L\mathcal{\sigma}_{\kappa_i}\}$ as 
\begin{equation}
    \rho=\frac{1}{2^L}
    \sum_{\kappa_1=0}^3\cdots
    \sum_{\kappa_L=0}^3
\left \langle \bigotimes_{i=1}^L\sigma_{\kappa_i} \right \rangle
\left(\bigotimes_{i=1}^L\sigma_{\kappa_i}\right),
\label{QST}
\end{equation}
where 
\begin{eqnarray} \sigma_0&=&\mathbbm{1}=\begin{pmatrix}
1 & 0 \\ 
0 & 1
\end{pmatrix},\ \sigma_1=X=\begin{pmatrix}
0 & 1 \\ 
1 & 0
\end{pmatrix},\notag\\
\sigma_2&=&Y=\begin{pmatrix}
0 & -i \\ 
i & 0
\end{pmatrix},\ \sigma_3=Z=\begin{pmatrix}
1 & 0 \\ 
0 & -1
\end{pmatrix}.
\end{eqnarray}
The basis elements are mutually orthogonal and span a $4^L$-dimensional linear space $\mathbb{D}$ (see Fig. \ref{space}).
The coefficient in Eq. (\ref{QST}) are expectation values on the state $\rho$, which ideally can be the results of experimental measurements.
Eq. (\ref{QST}) tells us that QST is experimentally challenging due to the the necessity (i) to measure different correlation functions whose number increase exponentially with respect to the system size, and (ii) to rotate the local spins (local quantization axes) site by site. 
QST has been limited to small systems (at most 8 qubits so far), leaving room for the development of scalable protocols for the analysis of phase transition phenomena.

One considerable advancement against the above issues is QST via compressed sensing \cite{CS1,CS2}, in which $\rho$ is reconstructed with the condition that $\rho$ is a low-rank state.
This is a relatively feasible protocol requiring $\mathcal{O}(r2^L(\log2^L)^2)$ measurement settings, where $r={\rm rank}\rho\ll2^L$, but the exponential factor remains. 
On the other hand, one may consider permutationally invariant (PI) QST \cite{PIQST1, PIQST2,PIQST3}, in which we recover the PI part of a density matrix (PIDM) $\mathit{\Pi}(\rho)$.
In PIQST, we only need $\mathcal{O}(L^2)$ measurement settings and only global rotations of the quantization axes. 
Besides, it is equivalent to full QST in case of PI states, defined by $\mathit{\Pi}(\rho)=\rho$, such as the GHZ states \cite{GHZ1,GHZ2}, the W states \cite{W1, W2, W3}, and the Dicke states \cite{Dicke}.

To obtain a scalable solution in the quantum simulations, we focus on the PIQST  owing to the drastically reduced experimental cost.
Although PIQST has been demonstrated experimentally \cite{PIQST1}, it has not yet been performed for large-size systems, especially for the application to the analysis of phase transition phenomena.
From the theoretical side, though we generally cannot reconstruct $\rho$ via PIQST, it has been reported that $\mathit{\Pi}(\rho)$ can encode the nonseparability of $\rho$ \cite{PIQST3}.
Also, any entanglement measure of $\mathit{\Pi}(\rho)$ is suggested to be the lower bound of that of $\rho$ \cite{PIQST3}.
However, none has been explored for the analysis of phase transition phenomena yet.
In this work, we study how the second-order Rényi entropy of $\mathit{\Pi}(\rho)$ reflects the properties of the ground state of the system when it undergoes a quantum phase transition.
We establish a scalable procedure, in which we directly compute the Rényi entropy of $\mathit{\Pi}(\rho)$ from only $\mathcal{O}(L^3)$ expectation values.
We calculate it for the ground state of the one-dimensional transverse-field Ising model \cite{TIM} using the density matrix renormalization group (DMRG) algorithm \cite{DMRG1,DMRG2,DMRG3}.

For the case of ferromagnetic (FM) interaction, the Rényi entropy is zero both in the limits of zero and infinite field because of the permutational invariance of the ground states.
It shows a broad peak around the quantum critical point (QCP), getting more pronounced for larger-size systems.
In the antiferromagnetic (AFM) case, the peak structure disappears and the Rényi entropy tends to diverge to infinity in the whole field range of the ordered phase.
By looking more closely at the scaling of the Rényi entropy versus system size,
we find that in AFM case the Rényi entropy grows almost linearly and its slope sharply changes around the QCP, while for the FM case the slopes around the QCP tend to take large values.
We also analyze the behavior of the Rényi entropy at low and high fields, according to the second-order perturbation theory.
We conjecture that the FM, AFM, and paramagnetic (PM) phases, which are the ground states of the transverse-field Ising model, are characterized by the different scaling law of the Rényi entropy of $\mathit{\Pi}(\rho)$.
We expect that origin of these behaviors will be clarified by the experimental measurement and theoretical analysis of larger-size systems, but we consider this study an important step in the analysis of phase transition phenomena utilizing quantum information. 

The remainder of this paper is organized as follows. 
In Sec. II, we review the PIQST and its performance.
In Sec. III, an analytical form of the second-order Rényi entropy of $\mathit{\Pi}(\rho)$ is given.
In Sec. IV, we show the results of the Rényi entropy of $\mathit{\Pi}(\rho)$ of the ground state of the transverse-field Ising chain across the quantum phase transition.
We also analyze its behavior in low- and high-field region according to second-order perturbation theory.
In Sec. V, we discuss the significant change of the Rényi entropy of PIDM around the QCP, and conjecture that the FM, AFM, and PM phases are characterized by its scaling law.
We also present an open question about the role of Rényi entropy of PIDM, and propose interesting future works to obtain deeper insights.
Section VI is devoted to the conclusion.

\section{permutationally invariant quantum state tomography}\label{sec:PIQST}
PIQST was introduced as a scalable protocol to reconstruct the density matrix of  a PI state, which has been mainly realized in photonic systems \cite{PIQST1, PIQST2}. 
In PIQST, we construct PIDM as
\begin{equation}
\mathit{\Pi}(\rho)=\frac{1}{L!}
\sum_{s\in \mathfrak{S}_L}\hat{\mathcal{P}}_s\rho\hat{\mathcal{P}}_s^\dagger,
\label{PIQST}
\end{equation}
where $\mathfrak{S}_L$ denotes the symmetric group of degree $L$ and $\hat{\mathcal{P}}_s$ is the unitary operator representing the permutation $s$ which rearranges $L$ qubits in a certain order: $\hat{\mathcal{P}}_s\ket{q_1q_2\cdots q_L}=\ket{q_{s(1)}q_{s(2)}\cdots q_{s(L)}}$.
We can easily confirm that $\mathit{\Pi}(\rho)$ has the fundamental properties of a density matrix: Hermicity, unit trace, and positive semi-definiteness.
An arbitrary PIDM exists in a linear space $\mathbb{P}$ (see Fig. \ref{space}), and  it can be expanded in the orthogonal PI basis $\{\mathcal{B}^{(k,l,m,n)}\}$ as
\begin{equation}
\mathit{\Pi}(\rho)=\frac{L!}{2^L}\sum_{ k+l+m+n=L}\frac{\braket{\mathcal{B}^{(k,l,m,n)}}}
{k!l!m!n!}\mathcal{B}^{(k,l,m,n)},
\label{PIQST2}
\end{equation}
where 
\begin{equation}
\mathcal{B}^{(k,l,m,n)}=\mathit{\Pi}\left(X^{\otimes k} \otimes Y^{\otimes l} \otimes Z^{\otimes m} \otimes \mathbbm{1}^{\otimes n}\right)
\label{PIQST3}
\end{equation}
and the sum $\sum_{ k+l+m+n=L}$ runs over all possible combinations of  non-negative integers, $k$, $l$, $m$, and $n$, which are the number of $X$, $Y$, $Z$, and $\mathbbm{1}$ in $\mathcal{B}^{(k,l,m,n)}$, respectively, subject to $k+l+m+n=L$.
By definition of the function $\mathit{\Pi}$, the PI basis elements which contain exactly same number of $X$, $Y$, $Z$, and $\mathbbm{1}$ are regarded as the same; e.g., $\mathit{\Pi}(X\otimes Y)=\mathit{\Pi}(Y\otimes X)$ for $L=2$, $\mathit{\Pi}(X\otimes X\otimes Y)=\mathit{\Pi}(X\otimes Y\otimes X)=\mathit{\Pi}(Y\otimes X\otimes X)$ for $L=3$, etc.
Naively this equality of correlation functions is the reason of the cost reduction in PIQST, which can be performed with merely $D_L$ measurement settings with
 \begin{equation}
     D_L\equiv\binom{L+2}{2}=\frac{L^2+3L+2}{2}.
 \end{equation} 
Thus the number of different expectation values to construct $\mathit{\Pi}(\rho)$ is
 \begin{equation}
     \sum_{n=0}^{L}D_{L-n}=\frac{L^3+6L^2+11L+6}{6}
 \end{equation}
in total. 
This is a powerful fact, since the exponential factor, which remains even in QST via compressed sensing \cite{CS1,CS2}, is absent here.
Moreover, there exists a linear transformation between the PI basis $\{\mathcal{B}^{(n)}_{\mu}\}_{\mu=1}^{D_{L-n}}$ and another basis $\{\mathcal{A}_{\nu}^{(n)}\}_{\nu=1}^{D_{L-n}}$, namely 
\begin{equation}
\mathcal{B}^{(n)}_{\mu}=\sum_{{\nu}=1}^{D_{L-n}}C_{\mu\nu}^{(n)}\mathcal{A}_{\nu}^{(n)},
\label{transform}
\end{equation}
where
\begin{equation}
\mathcal{A}_{\nu}^{(n)}=\mathit{\Pi}\left(A_{\nu}^{\otimes (L-n)} \otimes \mathbbm{1}^{\otimes n}\right)
\end{equation}
and $A_{\nu}=\alpha_{\nu}X+\beta_{\nu}Y+\gamma_{\nu}Z$, subject to $(\alpha_{\nu})^2+(\beta_{\nu})^2+(\gamma_{\nu})^2=1\ (\alpha_{\nu},\beta_{\nu},\gamma_{\nu}\in\mathbb{R})$, is a random operator acting on a local spin.
Here $\mathcal{B}_{\mu}^{(n)}=\mathcal{B}^{(k_{\mu},l_{\mu},m_{\mu},n)}$ and $k$, $l$, and $m$ are specified by the index $\mu$ (hence $k_{\mu}$, $l_{\mu}$, $m_{\mu}$).
The nontrivial single-site operators $A_\nu$ that compose $\mathcal{A}_{\nu}^{(n)}$ are identical, whereas those of $\mathcal{B}_{\mu}^{(n)}$ are generally different.
Through Eq. (\ref{transform}), we only need to measure the spins at all sites in the same direction to reconstruct $\mathit{\Pi}(\rho)$ in Eq. (\ref{PIQST2}).
Note that $\{\mathcal{A}_{\nu}^{(n)}\}$ are not required to be orthogonal as long as they are linearly independent, otherwise the linear space $\mathbb{P}$
spanned by $\{\mathcal{A}_{\nu}^{(n)}\}$ is not sufficiently large to express $\mathit{\Pi}(\rho)$.
In the experiments, for a four-photon Dicke state (PI state), it has been reported that $\mathit{\Pi}(\rho)$ gives high fidelity when taking an inhomogeneous distribution of the vectors $\mbox{\boldmath$a$}_{\nu}=(\alpha_{\nu},\beta_{\nu},\gamma_{\nu})^\top$ \cite{PIQST1};
This can be understood by taking the effect of noise into account.
The coefficients $C_{\mu\nu}^{(n)}$ are analytically determined by the sole information of the measurement directions $\{\mbox{\boldmath$a$}_{\nu}\}$. 
Let us define the vectors $\mbox{\boldmath$\mathcal{A}$}^{(n)}$ and $\mbox{\boldmath$\mathcal{B}$}^{(n)}$ as
\begin{eqnarray}
\mbox{\boldmath$\mathcal{A}$}^{(n)}&=&\left(\mathcal{A}_1^{(n)},\mathcal{A}_2^{(n)},\cdots,\mathcal{A}_{D_{L-n}}^{(n)}\right)^\top,\\
\mbox{\boldmath$\mathcal{B}$}^{(n)}&=&\left(\mathcal{B}_1^{(n)},\mathcal{B}_2^{(n)},\cdots,\mathcal{B}_{D_{L-n}}^{(n)}\right)^\top.
\end{eqnarray} 
Having fixed the number of identity matrices $n$, we decompose the matrix $\mathcal{A}_{\nu}^{(n)}$ into orthogonal basis elements $\{\mathcal{B}^{(n)}_{\mu}\}_{{\mu}=1}^{D_{L-n}}$, namely $\mbox{\boldmath$\mathcal{A}$}^{(n)}=(C^{(n)})^{-1}\mbox{\boldmath$\mathcal{B}$}^{(n)}$ or
$\mathcal{A}^{(n)}_\nu=\sum_{{\mu}=1}^{D_{L-n}}((C^{(n)})^{-1})_{\nu\mu}\mathcal{B}_{\mu}^{(n)}$, where $(C^{(n)})^{-1}$ is a $D_{L-n}$ by $D_{L-n}$ matrix in which the entries are given by
\begin{eqnarray}
((C^{(n)})^{-1})_{\nu\mu}&=&\frac{{\rm Tr}\left[\mathcal{A}^{(n)}_{\nu}\mathcal{B}^{(n)}_{\mu}\right]}{{\rm Tr}\left[\left(\mathcal{B}^{(n)}_{\mu}\right)^2\right]}\notag\\
&=&\frac{\left(2\alpha_{\nu}\right)^{k_{\mu}}\left(2\beta_{\nu}\right)^{l_{\mu}}\left(2\gamma_{\nu}\right)^{m_{\mu}}(L-n)!}{k_{\mu}!l_{\mu}!m_{\mu}!2^{L-n}}.\label{Cdd}
\end{eqnarray}
Thus we can calculate $\{\braket{\mathcal{B}^{(n)}_\mu}\}_{\mu=1}^{D_{L-n}}$ from $\{\braket{\mathcal{A}^{(n)}_\nu}\}_{\nu=1}^{D_{L-n}}$ using Eqs. (\ref{transform}) and (\ref{Cdd}) for each $n$. 
 
In this work, we focus on the second-order Rényi entropy \cite{Renyi} of PIDM, which is directly computed from a set of expectation values as shown in the next section.

\section{Second-order Rényi entropy of the permutationally invariant part of a density matrix}
Before studying the second-order Rényi entropy, let us review the purity, $P(\rho)={\rm Tr}(\rho^2)$.
Let us recall that it takes values between $1/2^L$ and $1$, in particular $P(\rho)=1$ for a pure states and $P(\rho)=1/2^L$ for a maximally mixed $L$-qubit state. 
In this work, we consider the purity of PIDM $P(\mathit{\Pi}(\rho))={\rm Tr}\left[\mathit{\Pi}(\rho)^2\right]$.
Since $\mathit{\Pi}(\rho)$ has the properties of a density matrix, $P(\mathit{\Pi}(\rho))$ also takes values between $1/2^L$ and $1$. 
$P(\mathit{\Pi}(\rho))$ can be expressed as follows (see also Appendix \ref{app_purityPIDM} for the derivation):
\begin{equation}
P(\mathit{\Pi}(\rho))=\frac{L!}{2^L}\sum_{k+l+m+n=L}\frac{\braket{\mathcal{B}^{(k,l,m,n)}}^2}{k!l!m!n!}.
\label{purity_PI}
\end{equation}

The second-order Rényi entropy is easy to calculate from  $P(\mathit{\Pi}(\rho))$ as $S_2(\mathit{\Pi}(\rho))=-\log P(\mathit{\Pi}(\rho))$ \cite{Renyi}, satisfying $0\leq S_2(\mathit{\Pi}(\rho))\leq L\log2$. 

\begin{figure}[t]
    \centering
    \includegraphics[width=8.4cm]{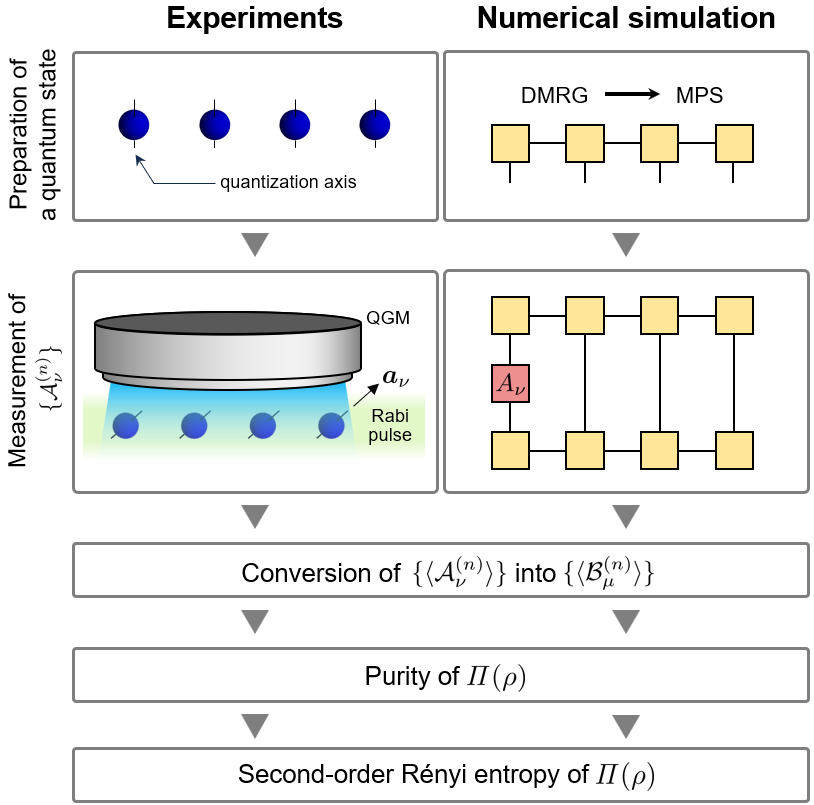}
    \caption{The protocols for evaluating the second-order Rényi entropy of PIDM in the experiments and in the numerical simulation (this work). In this work, we apply DMRG algorithm to obtain the ground state of a physical Hamiltonian and simulate the observations of $\mathcal{A}_\nu^{(n)}$ by calculating the expectation value of the matrix product operator $A_\nu$.
    The numerical calculations after the measurements are the same in the experimental protocol and on the simulations.}
   \label{protocol}
\end{figure}

We indicate a procedure for evaluating $S_2(\mathit{\Pi}(\rho))$ in experiments and numerical simulations (this work) in Fig. \ref{protocol}.
The numerical calculation steps to obtain $P(\mathit{\Pi}(\rho))$ are the following:
(i) we obtain the ground state of a physical Hamiltonian as a MPS \cite{MPS1, MPS2} using the DMRG algorithm \cite{DMRG1,DMRG2,DMRG3},
(ii) we calculate the expectation values $\{\braket{\mathcal{A}_\nu^{(n)}}\}$,
(iii) we compute the purity of PIDM by Eq. (\ref{purity_PI}), converting $\{\braket{\mathcal{A}_\nu^{(n)}}\}$ into $\{\braket{\mathcal{B}^{(n)}_\mu}\}$ via Eq. (\ref{transform}).

In optical-lattice experiments, a set of correlation functions $\{\braket{\mathcal{A}_\nu^{(n)}}\}$ can possibly be measured using quantum-gas microscope (QGM) \cite{QGM_Bakr,QGM_Cheuk,QGM_Haller,QGM_Miranda_1,QGM_Yamamoto,QGM_Miranda_2}, after a series of magnetic and optical operations that rotate the quantization axes to $\{\mbox{\boldmath$a$}_\nu\}$.
This experimental protocol is scalable because we can directly calculate the purity of PIDM from only $D_L$ measurement settings without constructing PIDM.

In the context of the analysis of quantum entanglement, the Rényi entanglement entropy $S_2(\rho_{\rm A})$, where $\rho_{\rm A}={\rm Tr_{\bar{A}}}\rho$ is a reduced density matrix, is often used to evaluate the amount of quantum entanglement between two complementary subsystems A and $\bar{\rm A}$ \cite{exp_ent1,exp_ent2,exp_ent3}.
In this work, we treat the Rényi entropy of the whole system, specifically regarding PIDM.
Naively, the latter acts as a measure of how a state $\rho$ is close to a PI state. 

\section{Model and Results}
\subsection{The transverse-field Ising chain}
As a case study, we compute the second-order Rényi entropy of the PIDM $S_2(\mathit{\Pi}(\rho))$ for the ground state of the transverse-field Ising chain \cite{TIM} with open boundary conditions, whose Hamiltonian is given by
\begin{eqnarray}
    \hat{\mathcal{H}}=J\sum_{i=1}^{L-1}\hat{Z}_i\hat{Z}_{i+1}-\mathit{\Gamma}\sum_{i=1}^L\hat{X}_i,
\label{TIM_Hamiltonian}
\end{eqnarray}
where $\hat{Z}_i$ and $\hat{X}_i$ are Pauli matrices acting on site $i$.
Here, we only deal with the case of even $L$.
In absence of $\mathit{\Gamma}$, the Hamiltonian is classical and features a trivial ground state.
For the case of FM interaction ($J<0$), the ground state is $\ket{\rm FM}=c_1\ket{\uparrow}^{\otimes L}+c_2\ket{\downarrow}^{\otimes L}$, and for the case of AFM interaction ($J>0$), the ground state is $\ket{{\rm AFM}}=c_1\ket{\uparrow\downarrow}^{\otimes \frac{L}{2}}+c_2\ket{\downarrow\uparrow}^{\otimes \frac{L}{2}}$, subject to $|c_1|^2+|c_2|^2=1$.
When an infinitesimal field is added, the ground state is symmetric, and we get $\ket{\rm FM}$ and $\ket{\rm AFM}$ with $c_1=c_2=1/\sqrt{2}$ \cite{TIM_dG}.
In the limit of infinite field (${\it \Gamma}\rightarrow\infty$), we can neglect the first term of Eq. (\ref{TIM_Hamiltonian}), yielding the PM phase which is given by the direct product of the eigenstates of $X$ as
\begin{equation}
     \ket{\rightarrow}^{\otimes L}=\frac{1}{2^{L/2}}
     \sum_{n=0}^L\sum_{s\in\mathfrak{S}_L}
     \frac{\hat{\mathcal{P}}_s\left(\ket{\uparrow}^{\otimes (L-n)}\otimes\ket{\downarrow}^{\otimes n}\right)}{(L-n)!n!},
     \label{ES_PX}
\end{equation}
where $\ket{\rightarrow}=(\ket{\uparrow}+\ket{\downarrow})/\sqrt{2}$.
In the intermediate field region, both terms contribute and a phase transition occurs at the QCP (${\it \Gamma}/J=1$).

\subsection{Behaviors of the second-order Rényi entropy of the permutationally invariant part for the ground state}
\begin{figure}[t]
    \centering
    \includegraphics[width=8.4cm]{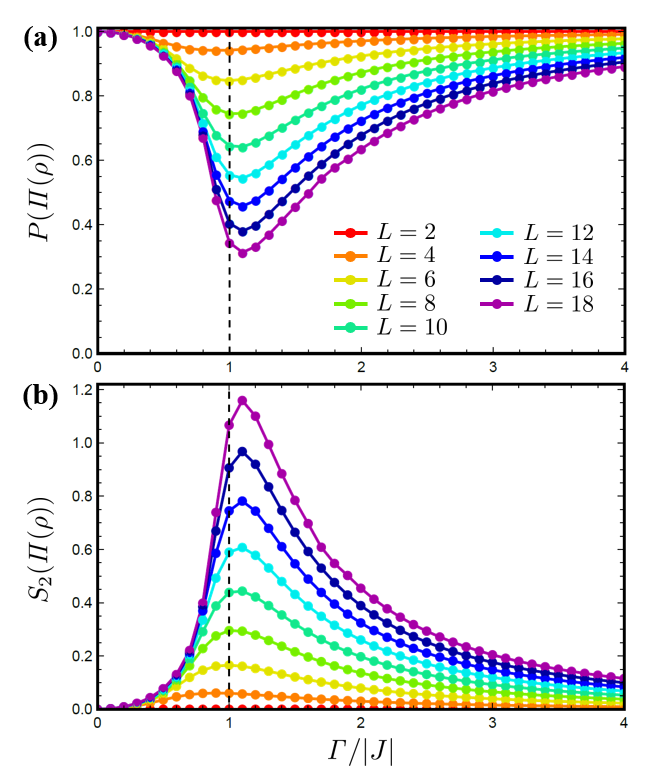}
    \caption{(a) The purity and (b) the second-order Rényi entropy of PIDM of the ground state of the FM transverse field Ising model as a function of the transverse-field for various system sizes.
The black dashed line represents the QCP ($\mathit{\Gamma}/|J|=1$).}
   \label{TIM_FM}
\end{figure}

\begin{figure}[t]
    \centering
    \includegraphics[width=8.4cm]{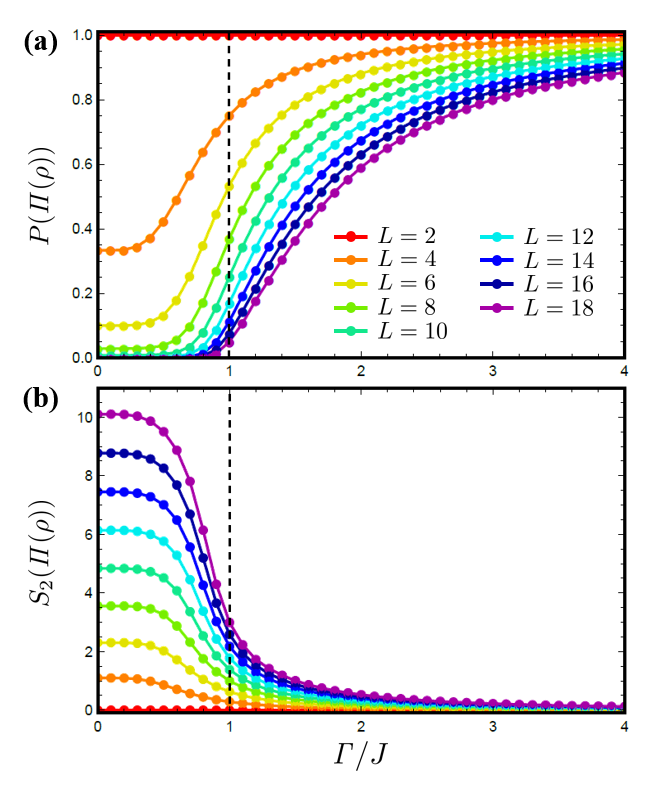}
    \caption{(a) The purity and (b) the second-order Rényi entropy of PIDM of the ground state of the AFM transverse-field Ising model as a function of the transverse field for various system sizes.
    The black dashed line represents the QCP ($\mathit{\Gamma}/J=1$). }. 

   \label{TIM_AFM}
\end{figure}

We show the plots of $P(\mathit{\Pi}(\rho))$ and $S_2(\mathit{\Pi}(\rho))$ as a function of $\mathit{\Gamma}/|J|$ in Figs. \ref{TIM_FM} and \ref{TIM_AFM} for the FM and AFM case, respectively. 
In the DMRG calculation, we take the full bond dimension (the maximum bond dimension of MPS $\chi_{\rm max}=2^{L/2}$) up to $L=10$;
we then set the cutoff of the maximum bond dimension $\chi_{\rm max}=16,10,4,4$ for $L=12,14,16,18$, respectively, due to the computational cost.
This truncation does not significantly affect the results because the purity of PIDM has essentially converged at $\chi_{\rm max}=4$ (see also Appendix \ref{app_chi} for the $\chi_{\rm max}$ dependence of $P(\mathit{\Pi}(\rho))$).

In the case of FM interaction, we obtain $S_2(\mathit{\Pi}(\rho))=0$ ($P(\mathit{\Pi}(\rho))=1$) at zero field.
For sufficiently large field, the values of $S_2(\mathit{\Pi}(\rho))$ approach zero ($P(\mathit{\Pi}(\rho))\rightarrow 1$).
These results are trivial because the ground states are PI states in these two limits.
For intermediate fields, there appears a broad peak around the QCP (black dashed line), which gets more pronounced for increasing system size.
In the case of AFM interaction, $S_2(\mathit{\Pi}(\rho))$ takes a finite value in the limit of zero field, caused by the non-PI ground states $\ket{{\rm AFM}}$.
In the limit of large field,  $S_2(\mathit{\Pi}(\rho))$ gets closer to zero, similar to the FM case.

It is well known that the QCP in the transverse-field Ising model can be characterized by various properties, such as the divergence of the entanglement entropy \cite{EE} and the maximum of next-nearest-neighbor entanglement \cite{nnnEE}.
Here a question arises: can $P(\mathit{\Pi}(\rho))$ or $S_2(\mathit{\Pi}(\rho))$ characterize the QCP in the thermodynamic limit?
We show the plots of $S_2(\mathit{\Pi}(\rho))$ as a function of $L$ for FM and AFM cases in Figs. \ref{TIM_fit}(a) and \ref{TIM_fit}(b), respectively.
(see Appendix \ref{app_chi} for the convergence of the numerical data as a function of $\chi_{\rm max}$).
We find that the data are well represented by fitting functions of the form
\begin{equation}
    f_{S_2}(L)=p_1L+p_2+\frac{p_3}{L}+p_4\log L,
    \label{fS}
\end{equation}
where $p_1$, $p_2$, $p_3$ and $p_4$ are fitting parameters (dashed lines in Fig. \ref{TIM_fit}).
We show the values of the fitting parameters in Figs. \ref{fitpara_FM} and \ref{fitpara_AFM}, determined by using the data between $L=2$ and $22$.
The error bars in Figs. \ref{fitpara_FM} and \ref{fitpara_AFM} represent plus/minus one standard deviation.
For the FM case, the behavior of the fitting functions are fundmentally different between FM and PM regions.
The largest slope of $S_2(\mathit{\Pi}(\rho))$ appears around the QCP, leading to a peak structure of $p_1$.
The peak is slightly shifted from the QCP, but we expect that it will eventually coincide with the QCP for larger-size systems.
For the AFM case, the slope of $S_2(\mathit{\Pi}(\rho))$ sharply decreases around the QCP for increasing ${\it \Gamma}$.

\begin{figure}[h]
    \centering
    \includegraphics[width=8.4cm]{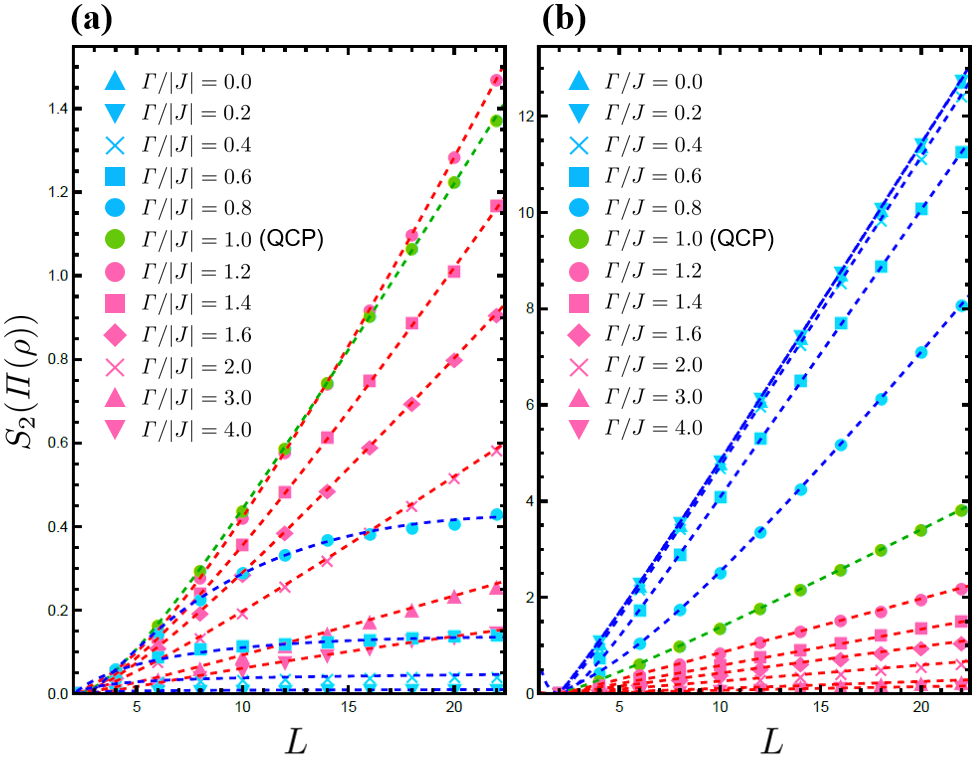}
    \caption{The second-order Rényi entropy of PIDM for the ground state of the transverse-field Ising chain as a function of $L$ in (a) FM case and (b) AFM cases. The dashed lines represent the fitting functions $f_{S_2}(L)$ defined by Eq. (\ref{fS}) with fitting parameters as shown in Figs. \ref{fitpara_FM} and \ref{fitpara_AFM}.}
   \label{TIM_fit}
\end{figure}

\subsection{Perturbative approach to low- and high-field regimes}

Let us consider the behavior of the second-order Rényi entropy under low (${\it \Gamma}/|J|\ll 1$) and high field ($|J|/{\it \Gamma}\ll 1$) according to the second-order perturbation theory (see also Appendix \ref{app_perturbation} for detailed calculations).
In the low-field region, the second-order Rényi entropy of the ground state is given by
\begin{equation}
       S_2(\mathit{\Pi}(\rho))\approx
   \left(\frac{\it \Gamma}{J}\right)^2\left(\frac{1}{4}-\frac{1}{2L}\right)\label{S_LF_FM}
   \end{equation}
for the FM case, and
\begin{eqnarray}
     S_2(\mathit{\Pi}(\rho))&\approx&L\log2-\frac{1}{2}\log L-\log\sqrt{2\pi}\notag\\
     &&\ \ \ \ \ \ \ \ \ -\left(\frac{\it \Gamma}{J}\right)^2\left(\frac{1}{4}+\frac{1}{2L}\right)\label{S_LF_AFM}
\end{eqnarray}
for the AFM case.
Thus, the fitting parameters in Eq. (\ref{fS}) are determined as
$(p_1,p_2,p_3,p_4)=(0,\frac{1}{4}(\frac{\it \Gamma}{J})^2,-\frac{1}{2}(\frac{\it \Gamma}{J})^2,0)$ and $(p_1,p_2,p_3,p_4)=(\log2,-\log\sqrt{2\pi}-\frac{1}{4}(\frac{\it \Gamma}{J})^2,-\frac{1}{2}(\frac{\it \Gamma}{J})^2,-\frac{1}{2})$ for FM and AFM cases, respectively (see also blue dashed lines in Figs. \ref{fitpara_FM} and \ref{fitpara_AFM}).
In the high-field region, the second-order Rényi entropy of the ground state is given by 
\begin{equation}
       S_2(\mathit{\Pi}(\rho))\approx
   \left(\frac{J}{\it \Gamma}\right)^2\left(\frac{L}{8}-\frac{3}{8}\right)\label{_HF}
\end{equation}
for the both FM and AFM cases, leading to fitting parameters $(p_1,p_2,p_3,p_4)=(\frac{1}{8}(\frac{J}{\it \Gamma})^2,-\frac{3}{8}(\frac{J}{\it \Gamma})^2,0,0)$ (see also red dashed lines in Figs. \ref{fitpara_FM} and \ref{fitpara_AFM}).

We have considered open boundary conditions so far.
With periodic boundary conditions, in the low-field region the Rényi entropy of PIDM is given by
\begin{equation}
       S_2(\mathit{\Pi}(\rho))\approx0
\end{equation}
 for the FM case and,
 \begin{equation}
     S_2(\mathit{\Pi}(\rho))\approx L\log2-\frac{1}{2}\log L-\log\sqrt{2\pi}
\end{equation}
for the AFM case, leading to fitting parameters $(p_1,p_2,p_3,p_4)=(0,0,0,0)$ and $(p_1,p_2,p_3,p_4)=(\log2,-\log\sqrt{2\pi},0,-\frac{1}{2})$, respectively.
Unlike open boundary conditions, the second-order term with respect to ${\it \Gamma}/J$ does not contribute to $S_2(\mathit{\Pi}(\rho))$, and the $L^{-1}$ term vanishes. 
In the high-field region, the Rényi entropy of PIDM is given by
\begin{equation}
       S_2(\mathit{\Pi}(\rho))\approx
   \left(\frac{J}{\it \Gamma}\right)^2\left(\frac{L}{8}-\frac{1}{4}\right)
\end{equation}
for the both FM and AFM cases, leading to $(p_1,p_2,p_3,p_4)=(\frac{1}{8}(\frac{J}{\it \Gamma})^2,-\frac{1}{4}(\frac{J}{\it \Gamma})^2,0,0)$.
The coefficient of $(J/{\it \Gamma})^2$ in $p_2$ is slightly different from the case with open boundary conditions, but the behavior is almost the same. 

\section{discussion}
\begin{figure}[t]
    \centering
    \includegraphics[width=7.4cm]{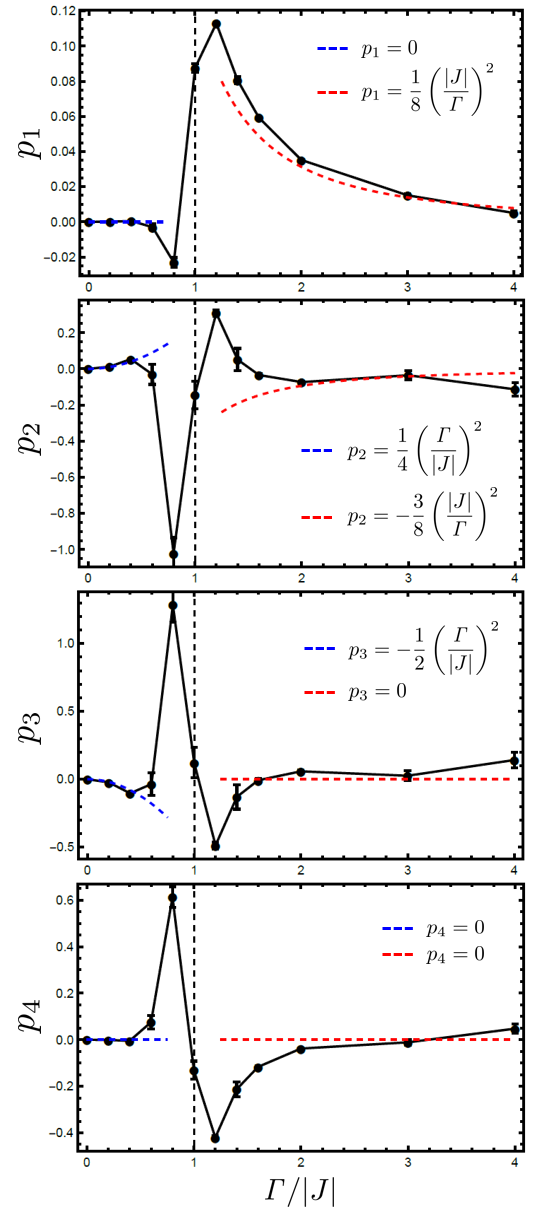}
    \caption{The fitting parameters as a function of ${\it \Gamma}/|J|$ for FM case. The blue and red dashed lines are analytical results by the second-order perturbation theory at low-field and high-field regions, respectively.
    Black dashed line represents the QCP.}
   \label{fitpara_FM}
\end{figure}

\begin{figure}[t]
    \centering
    \includegraphics[width=7.4cm]{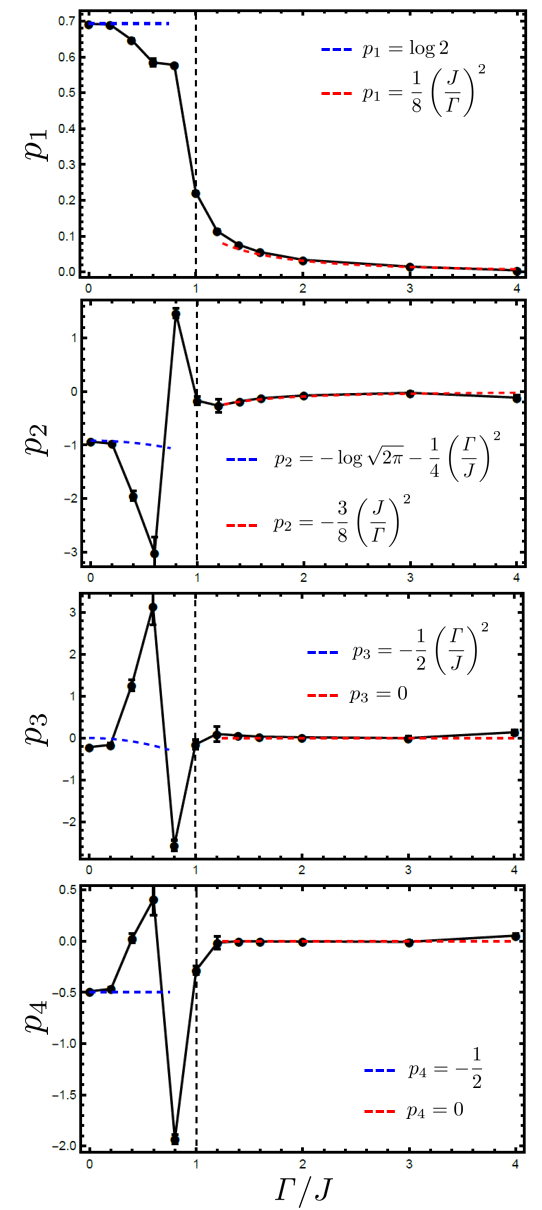}
    \caption{The fitting parameters as a function of ${\it \Gamma}/J$ for AFM case. The blue and red dashed lines are analytical results by the second-order perturbation theory at low-field and high-field regions, respectively.
    Black dashed line represents the QCP.}
   \label{fitpara_AFM}
\end{figure}

In this section, we summarize the behavior of  $S_2(\mathit{\Pi}(\rho))$ and discuss the possible application of $S_2(\mathit{\Pi}(\rho))$ for characterizing the phases of the system.
First, we can see that some properties of $S_2(\mathit{\Pi}(\rho))$ reflects the QCP.
For example, in Fig. \ref{TIM_fit} (a) the largest slope of $S_2(\mathit{\Pi}(\rho))$ appears around the QCP, leading to a peak structure of $p_1$.
In Fig. \ref{TIM_fit} (b), we can see that the slope of $S_2(\mathit{\Pi}(\rho))$ sharply decreases around the QCP for increasing ${\it \Gamma}$, bringing a sharp change of $p_1$.
Not only $p_1$, but also the other fitting parameters show significant changes, which can be associated with the criticality. 
The critical behavior of the fitting parameters in Figs. \ref{fitpara_FM} and \ref{fitpara_AFM} is slightly shifted from the QCP, but we ascribe this to the limitation of system size. 

Secondly, we conjecture that the finite-size scaling law of $S_2(\mathit{\Pi}(\rho))$ characterizes the FM, AFM, and PM phase.
For instance, in Fig. \ref{fitpara_FM}, $p_1$ is almost zero in the FM phase while it takes finite values in the PM phase.
In Fig. \ref{fitpara_AFM}, $p_4$ takes non-zero values in the AFM phase, but it almost vanishes in the PM phases.
From the results of the second-order perturbation theory, we expect that in the thermodynamic limit $S_2(\mathit{\Pi}(\rho))$ in the FM, AFM, and PM phases scales as
\begin{equation}
S_2(\mathit{\Pi}(\rho))=
\begin{cases} 
    \mathcal{O}(1) & \mbox{(FM)},\\
    p_1L+p_4\log L+\mathcal{O}(1)& \mbox{(AFM)},\\
    p_1L+\mathcal{O}(1)& \mbox{(PM)},\label{scale}
\end{cases}
\end{equation}respectively.

While in this work we investigate the behavior of $S_2(\mathit{\Pi}(\rho))$ across the quantum phase transition of the transverse-field Ising model, the question arises whether this behavior appears for a generic quantum phase transition.
In this respect, we observe that $S_2(\mathit{\Pi}(\rho))$ can be interpreted as a measure of how a state $\rho$ is close to a PI state, which in fact vanishes for an exactly PI state. 
It is then reasonable to expect that a quantum phase transition between two PI states will produce a peak structure in $S_2(\mathit{\Pi}(\rho))$  similar to Fig. \ref{TIM_FM} (b), whereas one between a PI and a non-PI state will bring about a sharp, step-like, change like in Fig. \ref{TIM_AFM} (b).
In order to get insights into the case of a transition between two non-PI state, the analysis of different models is required; however, we may speculate that another kind of step-like behavior occurs.
We consider it particularly interesting to study whether $S_2(\mathit{\Pi}(\rho))$ can distinguish between two states in which no standard order parameter can be defined, such as quantum spin liquids and topological phases.

By the definition Eq. (\ref{PIQST}), $\mathit{\Pi}(\rho)$ is the average over all possible site permutations of a state $\rho$.
We thus note that recovering $\mathit{\Pi}(\rho)$ instead of $\rho$ leads to a loss of information about the local properties of $\rho$. 
$S_2(\mathit{\Pi}(\rho))$ in some way quantifies this loss.

\section{conclusion}
We have studied the second-order Rényi entropy of PIDM $S_2(\mathit{\Pi}(\rho))$ for the ground state of the transverse-field Ising chain.
In the case of FM interaction, $S_2(\mathit{\Pi}(\rho))$ exhibits a broad peak structure around the QCP, which becomes more pronounced for increasing system size as shown in Fig. \ref{TIM_FM}.
In the AFM case, the peak structure disappears and $S_2(\mathit{\Pi}(\rho))$ tends to diverge to infinity in the whole range of the AFM phase  as shown in Fig. \ref{TIM_AFM}.
$S_2(\mathit{\Pi}(\rho))$ can be seen as a measure of permutational invariance of the (ground) state, which takes zero value for a PI state.
We have especially studied its behaviors as a function of system size, varying the intensity of the transverse field.
We have fitted the data of $S_2(\mathit{\Pi}(\rho))$ versus $L$ using a function containing linear, constant, $L^{-1}$ and logarithmic terms.
For the FM case, the fitting function approximately at the QCP has the largest slope.
We ascribe this slight shift to the limitation of system size and expect the fitting function at the QCP to be the upper bound of $S_2(\mathit{\Pi}(\rho))$ for sufficiently large $L$.
For the AFM case, we have found that the fitting parameters sharply change around the QCP.
We expect that the the fitting function at the QCP will be a boundary between AFM and PM phases.
We also analyze the behavior of $S_2(\mathit{\Pi}(\rho))$ under low and high fields according to the second-order perturbation theory.
We conjecture that the FM, AFM, and PM phases are distinguished by the scaling law of $S_2(\mathit{\Pi}(\rho))$ as shown in Eq. (\ref{scale}).

Lastly, let us emphasize that, while our simulations are practically limited to $L=22$ by the explicit calculation of ${\it \Pi}(\rho)$ from $\rho$, the prospective experimental protocols and subsequent calculation of purity via Eqs. \ref{transform}, \ref{Cdd}, and \ref{purity_PI} would be in principle scalable to a much larger system size, being dependent upon the measurement of only $\mathcal{O}(L^3)$ correlations (corresponding to only $\mathcal{O}(L^2)$ experimental settings).
We hope that future efforts will clarify the properties of $S_2(\mathit{\Pi}(\rho))$ through both experimental measurements and theoretical analysis of other physical models, possibly acquiring some insights into the connection with symmetry breaking, conformal field theory, topological orders, etc.

\begin{acknowledgments}
We thank T. Shimokawa, D. Kagamihara, S. Goto, R. Kaneko, and I. Danshita for useful discussions. 
This work was supported by JSPS KAKENHI 
Grant Nos. 
23KJ1867 (Y.M.), 
18K03525, 21H05185, 23K25830 and 24K06890 (D.Y.), 
22H01171 (N.F., D.Y.), 
JST PRESTO Grant No. JPMJPR2118, Japan (D.Y.),
and Aoyama Gakuin University Research Institute “Early Eagle” grant program for promotion of research by early career researchers.
\end{acknowledgments}

\appendix
\section{ANALYTICAL CALCULATION OF THE PURITY OF THE PIDM EXPRESSED IN THE $\mathcal{B}^{(k,l,m,n)}$ BASIS}
\label{app_purityPIDM}

In this appendix, we briefly give the derivation of Eq. (\ref{purity_PI}).
First, let us consider the trace of $(\mathcal{B}^{(k,l,m,n)})^2=\mathit{\Pi}(X^{\otimes k} \otimes Y^{\otimes l} \otimes Z^{\otimes m} \otimes \mathbbm{1}^{\otimes n})^2$.
\begin{widetext}
\begin{eqnarray}
    && {\rm Tr}\left[(\mathcal{B}^{(k,l,m,n)})^2\right]\notag\\
     &=&\frac{1}{\left(\frac{L!}{k!l!m!n!}\right)^2}{\rm Tr}\left[\left(\sum_{s\in\tilde{\mathfrak{S}}_L}\hat{\mathcal{P}}_s(X^{\otimes k} \otimes Y^{\otimes l} \otimes Z^{\otimes m} \otimes \mathbbm{1}^{\otimes n})\hat{\mathcal{P}}_s^\dagger\right)\left(\sum_{s'\in\tilde{\mathfrak{S}}_L}\hat{\mathcal{P}}_{s'}(X^{\otimes k} \otimes Y^{\otimes l} \otimes Z^{\otimes m} \otimes \mathbbm{1}^{\otimes n})\hat{\mathcal{P}}_{s'}^\dagger\right)\right]\notag\\
     &=&\frac{1}{\left(\frac{L!}{k!l!m!n!}\right)^2}{\rm Tr}\left[\sum_{s\in\tilde{\mathfrak{S}}_L}
     \hat{\mathcal{P}}_s\left((X^2)^{\otimes k} \otimes (Y^2)^{\otimes l} \otimes (Z^2)^{\otimes m} \otimes (\mathbbm{1}^2)^{\otimes n}\right)\hat{\mathcal{P}}_s^\dagger\right]\notag\\
     &=&\frac{1}{\left(\frac{L!}{k!l!m!n!}\right)^2}\sum_{s\in\tilde{\mathfrak{S}}_L}{\rm Tr}\left[
     \hat{\mathcal{P}}_s \left(\mathbbm{1}^{\otimes L}\right)\hat{\mathcal{P}}_s^\dagger\right]
     \ =\ \frac{1}{\left(\frac{L!}{k!l!m!n!}\right)^2}\times\frac{L!}{k!l!m!n!}\times ({\rm Tr}\mathbbm{1})^L
     \ =\ \frac{k!l!m!n!}{L!}2^L,
\end{eqnarray}
\end{widetext}
where $\sum_{\sigma\in\tilde{\mathfrak{S}}_L}$ denotes the summation over all the non-trivial permutations.
Subsequently, we calculate $\mathit{\Pi}(\rho)^2$ as follows:
\begin{widetext}
\begin{eqnarray}
\mathit{\Pi}(\rho)^2&=&\left\{\sum_{k+l+m+n=L}\frac{\braket{\mathit{\Pi}\left(X^{\otimes k} \otimes Y^{\otimes l} \otimes Z^{\otimes m} \otimes \mathbbm{1}^{\otimes n}\right)}}
{{\rm Tr}\left[(\mathit{\Pi}\left(X^{\otimes k} \otimes Y^{\otimes l} \otimes Z^{\otimes m} \otimes \mathbbm{1}^{\otimes n}\right))^2\right]}\mathit{\Pi}\left(X^{\otimes k} \otimes Y^{\otimes l} \otimes Z^{\otimes m} \otimes \mathbbm{1}^{\otimes n}\right)\right\}\notag\\
&&\times\left\{\sum_{k'+l'+m'+n'=L}\frac{\braket{\mathit{\Pi}(X^{\otimes k'} \otimes Y^{\otimes l'} \otimes Z^{\otimes m'} \otimes \mathbbm{1}^{\otimes n'})}}
{{\rm Tr}\left[(\mathit{\Pi}\left(X^{\otimes k'} \otimes Y^{\otimes l'} \otimes Z^{\otimes m'} \otimes \mathbbm{1}^{\otimes n'}\right))^2\right]}\mathit{\Pi}(X^{\otimes k'} \otimes Y^{\otimes l'} \otimes Z^{\otimes m'} \otimes \mathbbm{1}^{\otimes n'})\right\}.
\end{eqnarray}
\end{widetext}
There are $\left(\sum_{n=0}^LD_{L-n}\right)\times\left(\sum_{n'=0}^LD_{L-n'}\right)$ terms in total, but after taking the trace most of them do not contribute and only $\sum_{n=0}^LD_{L-n}$ terms are left due to the orthogonality
\begin{eqnarray}
    &&{\rm Tr}\left[\mathcal{B}^{(k,l,m,n)}\mathcal{B}^{(k',l',m',n')}\right]\notag\\
    &=&
    {\rm Tr}\left[(\mathcal{B}^{(k,l,m,n)})^2\right]\delta_{k,k'}\delta_{l,l'}\delta_{m,m'}\delta_{n,n'}\notag\\
    &=&\frac{k!l!m!n!}{L!}2^L\delta_{k,k'}\delta_{l,l'}\delta_{m,m'}\delta_{n,n'}.
\end{eqnarray}
Eventually we obtain the following formula:
\begin{eqnarray}
{\rm Tr}\left[\mathit{\Pi}(\rho)^2\right]&=&
\sum_{k+l+m+n=L}\frac{\braket{\mathcal{B}^{(k,l,m,n)}}^2}{{\rm Tr}\left[(\mathcal{B}^{(k,l,m,n)})^2\right]}\notag\\
&=&\frac{L!}{2^L}\sum_{k+l+m+n=L}\frac{\braket{\mathcal{B}^{(k,l,m,n)}}^2}{k!l!m!n!}.
\end{eqnarray}

\section{SECOND-ORDER PERTURBATION THEORY}
\label{app_perturbation}
In this appendix, we show the analytical calculation of $S_2(\mathit{\Pi}(\rho))$ at low and high field, according to the second-order perturbation theory.
We focus on the transverse-field Ising chain with open boundary conditions for even $L$.

\subsection{High-field regime}
Let us consider the Hamiltonian of the transverse-field Ising chain with a perturbation parameter $\lambda=J/{\it \Gamma}$ as
\begin{equation}
    \hat{\mathcal{H}}/{\it \Gamma}=
    \hat{\mathcal{H}}^{(0)}+\lambda
    \hat{\mathcal{H}}',
\end{equation}
where
\begin{eqnarray}
    \hat{\mathcal{H}}^{(0)}&=&-\sum_{i=1}^L\hat{X}_i,\\
    \hat{\mathcal{H}}'&=&\sum_{i=1}^{L-1}\hat{Z}_i\hat{Z}_{i+1}
\end{eqnarray}
are the unperturbed Hamiltonian and the perturbation, respectively.
$\hat{\mathcal{H}}^{(0)}$ provides the nondegenerate ground state $\ket{\rightarrow}^{\otimes L}$ , where $\ket{\rightarrow}=(\ket{\uparrow}+\ket{\downarrow})/\sqrt{2}$ , and the ground state of  $\hat{\mathcal{H}}/{\it \Gamma}$, denoted $\ket{\rm HF}$, is given by
\begin{equation}
 \ket{\rm HF}\simeq\frac{1}{\mathcal{N}}\left(\ket{\rightarrow}^{\otimes L}-\frac{\lambda}{4}\sum_{m=0}^{L-2}\ket{\psi_m}\right),   
\end{equation}
where $\ket{\psi_m}=\ket{\rightarrow}^{\otimes m}\otimes\ket{\leftarrow}^{\otimes 2}\otimes\ket{\rightarrow}^{\otimes (L-m-2)}$ with $\ket{\leftarrow}=\hat{Z}\ket{\rightarrow}=(\ket{\uparrow}-\ket{\downarrow})/\sqrt{2}$, and $\mathcal{N}=\left(1+\frac{1}{16}\lambda^2(L-1)\right)^{1/2}$ is a normalization factor.
Note that the $\mathcal{O}(\lambda^2)$ term of $\ket{\rm HF}$ is zero.
PIDM of $\ket{\rm HF}\bra{\rm HF}$ is given by 
\begin{widetext}
\begin{equation}
    {\it\Pi}(\ket{\rm HF}\bra{\rm HF})=\frac{1}{\mathcal{N}^2}\frac{1}{L!}\left[L!\ket{\rightarrow}^{\otimes L}\bra{\rightarrow}^{\otimes L}
    -\frac{\lambda}{4}\sum_{s\in \mathfrak{S}_L}\sum_{m=0}^{L-2}
    \left(\ket{\rightarrow}^{\otimes L}\bra{\psi_m}\hat{\mathcal{P}}_s^\dagger+{\rm H.c.}\right)
    \right],
\end{equation}
 \end{widetext}
 which is sufficient to include up to $\mathcal{O}(\lambda^2)$ terms in the second-order Rényi entropy of PIDM.
 The purity and second-order Rényi entropy of PIDM up to second order in the perturbation parameter are calculated as follows:
\begin{widetext}
\begin{eqnarray}
    P({\it\Pi}(\ket{\rm HF}\bra{\rm HF}))&=&\frac{1}{\mathcal{N}^4}\frac{1}{(L!)^2}\left[(L!)^2+2\times
    \frac{\lambda^2}{16}\sum_{s,s'\in \mathfrak{S}_L}\sum_{m,m'=0}^{L-2}
    {\rm Tr}\left(\ket{\rightarrow}^{\otimes L}\bra{\psi_m}\hat{\mathcal{P}}_s^\dagger\hat{\mathcal{P}}_{s'}\ket{\psi_{m'}}\bra{\rightarrow}^{\otimes L}\right)
    \right]\notag\\
    &=&\frac{1}{\mathcal{N}^4}\frac{1}{(L!)^2}\left[(L!)^2+2\times
    \frac{\lambda^2}{16}(L-2)!2!L!(L-1)^2
    \right]\notag\\
     &\simeq&1-\lambda^2\left(\frac{L}{8}-\frac{3}{8}\right),\\
    S_2({\it\Pi}(\ket{\rm HF}\bra{\rm HF}))&=&-\log\left[1-\lambda^2\left(\frac{L}{8}-\frac{3}{8}\right)\right]\notag\\
    &\simeq&\lambda^2\left(\frac{L}{8}-\frac{3}{8}\right).
    \end{eqnarray}
\end{widetext}

 \subsection{Low-field regime}
In order to analyze the behavior of the Rényi entropy of PIDM in the low-field region, we consider the following Hamiltonian 
\begin{equation}
\hat{\mathcal{H}}/|J|=\hat{\mathcal{H}}^{(0)}+\lambda
    \hat{\mathcal{H}}',\label{H_LF}    
\end{equation}
where
\begin{eqnarray}
    \hat{\mathcal{H}}^{(0)}&=&{\rm sgn}(J)\sum_{i=1}^{L-1}\hat{Z}_i\hat{Z}_{i+1},\label{H0_LF}\\
     \hat{\mathcal{H}}'&=&-\sum_{i=1}^L\hat{X}_i.
\end{eqnarray}
Here we redefine the perturbation parameter as $\lambda={\it \Gamma}/|J|$.
Let us begin with the FM case ($J<0$).
The ground state of $\hat{\mathcal{H}}^{(0)}$ is two-fold degenerate, namely $\ket{\uparrow}^{\otimes L}$ and $\ket{\downarrow}^{\otimes L}$, and the $L$th-order perturbation gives a small energy split between symmetrized and antisymmetrized combinations, which favors $(\ket{\uparrow}^{\otimes L}+\ket{\downarrow}^{\otimes L})/\sqrt{2}$.
Here we define
\begin{eqnarray}
\ket{\psi_m}&=&\ket{\uparrow}^{\otimes m}\otimes\ket{\downarrow}\otimes\ket{\uparrow}^{\otimes (L-m-1)},
\end{eqnarray}
\begin{eqnarray}
    \ket{\phi_m}&=&\ket{\downarrow}^{\otimes m}\otimes\ket{\uparrow}\otimes\ket{\downarrow}^{\otimes (L-m-1)}.
\end{eqnarray}
The ground state of  $\hat{\mathcal{H}}/|J|$, denoted $\ket{\rm LF_{FM}}$, is given by
\begin{widetext}
\begin{equation}
 \ket{\rm LF_{FM}}\simeq\frac{1}{\mathcal{N}}\left(\frac{\ket{\uparrow}^{\otimes L}+\ket{\downarrow}^{\otimes L}}{\sqrt{2}}-\frac{\lambda}{2}\left(\frac{\ket{\psi_0}+\ket{\phi_0}}{\sqrt{2}}+
 \frac{\ket{\psi_{L-1}}+\ket{\phi_{L-1}}}{\sqrt{2}}\right)
 -\frac{\lambda}{4}\sum_{m=1}^{L-2}\frac{\ket{\psi_m}+\ket{\phi_m}}{\sqrt{2}}\right),   
\end{equation}
\end{widetext}
where $\mathcal{N}=\left[1+\lambda^2\left(\frac{1}{2}-\frac{1}{16}(L-2)\right)\right]^{1/2}$ is a normalization factor.
Note that the $\mathcal{O}(\lambda^2)$ term of $\ket{\rm LF_{FM}}$ is zero.
PIDM of $\ket{\rm LF_{FM}}\bra{\rm LF_{FM}}$ is given by 
\begin{widetext}
\begin{equation}
    {\it\Pi}(\ket{\rm LF_{FM}}\bra{\rm LF_{FM}})=\frac{1}{\mathcal{N}^2}\frac{1}{L!}(A+(B+C+{\rm H.c.})),
\end{equation}
where
\begin{eqnarray}
    A&=&\frac{L!}{2}\left(\ket{\uparrow}^{\otimes L}+\ket{\downarrow}^{\otimes L}\right)\left(\bra{\uparrow}^{\otimes L}+\bra{\downarrow}^{\otimes L}\right),\\
B&=&-\frac{\lambda}{4}\sum_{s\in \mathfrak{S}_L}\left(
    \ket{\uparrow}^{\otimes L}+\ket{\downarrow}^{\otimes L}\right)
    \left(\bra{\psi_0}+\bra{\phi_0}+\bra{\psi_{L-1}}+\bra{\phi_{L-1}}\right)\hat{\mathcal{P}}^\dagger_s,\\
C&=&-\frac{\lambda}{8}\sum_{s\in \mathfrak{S}_L}\sum_{m=1}^{L-2}\left(
    \ket{\uparrow}^{\otimes L}+\ket{\downarrow}^{\otimes L}\right)
\left(\bra{\psi_m}+\bra{\phi_m}\right)\hat{\mathcal{P}}^\dagger_s.
\end{eqnarray}
 \end{widetext}
 The purity and second-order Rényi entropy of PIDM within second-order perturbation theory are calculated as follows:
\begin{widetext}
\begin{eqnarray}
    &&P({\it\Pi}(\ket{\rm LF_{FM}}\bra{\rm LF_{FM}}))\notag\\
    &=&\frac{1}{\mathcal{N}^4}\frac{1}{(L!)^2}{\rm Tr}(A^2+2BB^\dagger+2BC^\dagger+2CB^\dagger+2CC^\dagger)\notag\\
&=&\frac{1}{\mathcal{N}^4}\frac{1}{(L!)^2}\left\{
    (L!)^2+2\times\frac{\lambda^2}{4}4(L-1)!L!+4\times\frac{\lambda^2}{8}2(L-2)(L-1)!L!+2\times\frac{\lambda^2}{16}(L-2)^2(L-1)!L!\right\}\notag\\
    &\simeq&1-\lambda^2\left(\frac{1}{4}-\frac{1}{2L}\right),\\
     &&S_2({\it\Pi}(\ket{\rm LF_{FM}}\bra{\rm LF_{FM}}))=-\log\left[1-\lambda^2\left(\frac{1}{4}-\frac{1}{2L}\right)\right]\simeq\lambda^2\left(\frac{1}{4}-\frac{1}{2L}\right).
    \end{eqnarray}
\end{widetext}

For the AFM case ($J>0$), the ground state of the unperturbed Hamiltonian Eq. (\ref{H0_LF}) is two-fold degenerate, namely $\ket{\uparrow\downarrow}^{\otimes \frac{L}{2}}$ and $\ket{\downarrow\uparrow}^{\otimes \frac{L}{2}}$, and the $L$th-order perturbation gives a small energy split between symmetrized and antisymmetrized states of them, selecting $(\ket{\uparrow\downarrow}^{\otimes \frac{L}{2}}+\ket{\downarrow\uparrow}^{\otimes \frac{L}{2}})/\sqrt{2}$.
Here let us define $\ket{\psi_m^{(\uparrow)}},\ket{\psi_m^{(\downarrow)}},\ket{\phi_m^{(\uparrow)}}$, and $\ket{\phi_m^{(\downarrow)}}$ as
\begin{eqnarray}
\ket{\psi_m^{(\uparrow)}}&=&\ket{\uparrow\downarrow}^{\otimes m}\otimes\ket{\uparrow\uparrow}\otimes\ket{\uparrow\downarrow}^{\otimes (\frac{L}{2}-m-1)}, \\
\ket{\psi_m^{(\downarrow)}}&=&\ket{\uparrow\downarrow}^{\otimes m}\otimes\ket{\downarrow\downarrow}\otimes\ket{\uparrow\downarrow}^{\otimes (\frac{L}{2}-m-1)},
\end{eqnarray}
\begin{eqnarray}
\ket{\phi_m^{(\uparrow)}}&=&\ket{\downarrow\uparrow}^{\otimes m}\otimes\ket{\uparrow\uparrow}\otimes\ket{\downarrow\uparrow}^{\otimes (\frac{L}{2}-m-1)}, \\
\ket{\phi_m^{(\downarrow)}}&=&\ket{\downarrow\uparrow}^{\otimes m}\otimes\ket{\downarrow\downarrow}\otimes\ket{\downarrow\uparrow}^{\otimes (\frac{L}{2}-m-1)}.
\end{eqnarray} 
The ground state of Eq. (\ref{H_LF}), denoted $\ket{\rm LF_{AFM}}$, is given by
\begin{widetext}
\begin{eqnarray}
 \ket{\rm LF_{AFM}}&\simeq&\frac{1}{\mathcal{N}}\left[\frac{\ket{\uparrow\downarrow}^{\otimes \frac{L}{2}}+\ket{\downarrow\uparrow}^{\otimes \frac{L}{2}}}{\sqrt{2}}+\frac{\lambda}{4\sqrt{2}}\left(\sum_{m=0}^{\frac{L}{2}-2}\left(\ket{\psi^{(\uparrow)}_m}+\ket{\phi^{(\downarrow)}_m}\right)+
 \sum_{m=1}^{\frac{L}{2}-1}\left(\ket{\psi^{(\downarrow)}_m}+\ket{\phi^{(\uparrow)}_m}\right)\right)\right.\notag\\
 &&\ \ \ \ \ \ \ \ \ \ \ \ \ \ \ \ \ \ \ \ \ \ \ \ \ \ \ \ \ \ \ \ \ \ \ \ \ \ \ \ \ \left.\textcolor{white}{\frac{1}{1}}+\frac{\lambda}{2\sqrt{2}}\left(\ket{\psi^{(\uparrow)}_{\frac{L}{2}-1}}+\ket{\phi^{(\downarrow)}_{\frac{L}{2}-1}}+
\ket{\psi^{(\downarrow)}_0}+\ket{\phi^{(\uparrow)}_0}\right)\right],
\end{eqnarray}
\end{widetext}
where $\mathcal{N}=\left[1+\lambda^2\left(\frac{L}{16}+\frac{3}{8}\right)\right]^{1/2}$ is a normalization factor.
Note that $\mathcal{O}(\lambda^2)$ term of $\ket{\rm LF_{AFM}}$ is zero.
PIDM of $\ket{\rm LF_{AFM}}\bra{\rm LF_{AFM}}$ is given by

\begin{widetext}
\begin{equation}
    {\it\Pi}(\ket{\rm LF_{AFM}}\bra{\rm LF_{AFM}})=\frac{1}{\mathcal{N}^2}\frac{1}{L!}(A+(B+C+{\rm H.c.})),
\end{equation}
where
\begin{eqnarray}
    A&=&\frac{1}{2}\sum_{s\in \mathfrak{S}_L}\hat{\mathcal{P}}_s\left(\ket{\uparrow\downarrow}^{\otimes \frac{L}{2}}+\ket{\downarrow\uparrow}^{\otimes \frac{L}{2}}\right)\left(
    \bra{\uparrow\downarrow}^{\otimes \frac{L}{2}}+\bra{\downarrow\uparrow}^{\otimes \frac{L}{2}}\right)\hat{\mathcal{P}}_s^\dagger,\\
    B&=&\frac{\lambda}{8}\sum_{s\in \mathfrak{S}_L}\left[\hat{\mathcal{P}}_s
    \left(\ket{\uparrow\downarrow}^{\otimes \frac{L}{2}}+\ket{\downarrow\uparrow}^{\otimes \frac{L}{2}}\right)
    \left(\sum_{m=0}^{\frac{L}{2}-2}\left(\bra{\psi^{(\uparrow)}_m}+\bra{\phi^{(\downarrow)}_m}\right)+
 \sum_{m=1}^{\frac{L}{2}-1}\left(\bra{\psi^{(\downarrow)}_m}+\bra{\phi^{(\uparrow)}_m}\right)\right)\hat{\mathcal{P}}_s^\dagger\right],\\
    C&=&\frac{\lambda}{4}\sum_{s\in \mathfrak{S}_L}\left[\hat{\mathcal{P}}_s
    \left(\ket{\uparrow\downarrow}^{\otimes \frac{L}{2}}+\ket{\downarrow\uparrow}^{\otimes \frac{L}{2}}\right)
    \left(\bra{\psi^{(\uparrow)}_{\frac{L}{2}-1}}+\bra{\phi^{(\downarrow)}_{\frac{L}{2}-1}}+
\bra{\psi^{(\downarrow)}_0}+\bra{\phi^{(\uparrow)}_0}\right)\hat{\mathcal{P}}_s^\dagger\right].
\end{eqnarray}
 \end{widetext}
 The purity and second-order Rényi entropy of PIDM within second-order perturbation theory are calculated as follows:
\begin{widetext}
\begin{eqnarray}
    &&P({\it\Pi}(\ket{\rm LF_{AFM}}\bra{\rm LF_{AFM}}))\notag\\
    &=&\frac{1}{\mathcal{N}^4}\frac{1}{(L!)^2}{\rm Tr}(A^2+2BB^\dagger+2BC^\dagger+2CB^\dagger+2CC^\dagger)\notag\\
&=&\frac{1}{\mathcal{N}^4}\frac{1}{(L!)^2}\left\{
    2\left(\frac{L}{2}\right)!\left(\frac{L}{2}\right)!L!
    +2\times\frac{\lambda^2}{4}\left(\frac{L}{2}\right)!\left(\frac{L}{2}-1\right)!L!\left(\frac{L}{2}-1\right)^2\right.\notag\\
    &&\left.\ \ \ \ \ \ \ \ \ \ \ \ \ \ \ \ \ \ \ \ \ \ \ \ \ \ \ \ \ \ \ \ \ \ \ \ \  
       +4\times\frac{\lambda^2}{2}\left(\frac{L}{2}\right)!\left(\frac{L}{2}-1\right)!L!\left(\frac{L}{2}-1\right)+2\times\lambda^2\left(\frac{L}{2}\right)!\left(\frac{L}{2}-1\right)!L!\right\}\notag\\
      &\simeq&\frac{2\left(\frac{L}{2}\right)!\left(\frac{L}{2}\right)!}{L!}+\lambda^2\left(-\frac{2\left(\frac{L}{2}\right)!\left(\frac{L}{2}\right)!}{L!}\left(\frac{L}{8}+\frac{3}{4}\right)+\frac{\left(\frac{L}{2}\right)!\left(\frac{L}{2}-1\right)!}{L!}\left(\frac{L^2}{8}+\frac{L}{2}+\frac{1}{2}\right)\right),\\
&&S_2({\it\Pi}(\ket{\rm LF_{AFM}}\bra{\rm LF_{AFM}}))\notag\\
    &=&-\log\left[\frac{2\left(\frac{L}{2}\right)!\left(\frac{L}{2}\right)!}{L!}+\lambda^2\left(-\frac{2\left(\frac{L}{2}\right)!\left(\frac{L}{2}\right)!}{L!}\left(\frac{L}{8}+\frac{3}{4}\right)+\frac{\left(\frac{L}{2}\right)!\left(\frac{L}{2}-1\right)!}{L!}\left(\frac{L^2}{8}+\frac{L}{2}+\frac{1}{2}\right)\right)\right]\notag\\
    &\simeq&L\log2-\frac{1}{2}\log L-\log\sqrt{2\pi}-\lambda^2\left(\frac{1}{4}+\frac{1}{2L}\right)\label{S_LF_AFM_append}
    \end{eqnarray}
\end{widetext}
Here we have used the Stirling’s formula to obtain Eqs. \ref{S_LF_AFM_append}.

\section{MAXIMUM BOND DIMENSION DEPENDENCE OF  THE PURITY OF PIDM}
\label{app_chi}
\begin{figure*}[t]
 \begin{center}
  \includegraphics[width=15cm]{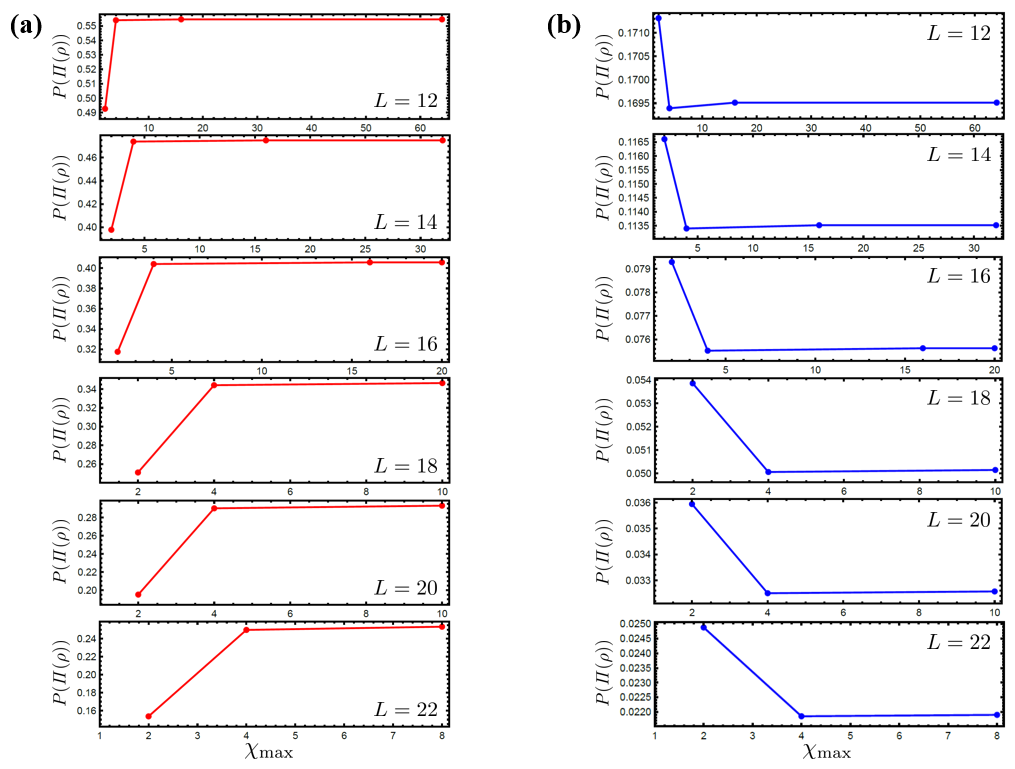}
  \label{TIM_QCP_chi}
  \caption{The maximum bond dimension $\chi_{\rm max}$ dependence of the purity of PIDM for the ground state of (a) the FM and (b) the AFM transverse-field Ising models at the QCP.}
 \end{center}
\end{figure*}

In this appendix, we show the plots of the purity of PIDM for the ground state of the transverse-field Ising chain as a function of the maximum bond dimension of MPS $\chi_{\rm max}$ in Fig. \ref{TIM_QCP_chi}. 
We can see that the purity sufficiently converges at $\chi_{\rm max}\simeq4$.

\end{document}